# Comment on "Proton and kaon production in Au+Au collisions at $\sqrt{s_{NN}}$ = 3 GeV"

V. A. Kizka[1,2]

[1]*Institute for Safety Problems of Nuclear Power Plants, Chornobyl, Ukraine*

[2]*V.N.Karazin Kharkiv National University, Kharkiv, Ukraine*

**Abstract:** Shuang-Jie Liu et al. (arXiv:2604.27678) reported that the momentum-dependent nuclear mean field (MDI) with $K_0$ = 230 MeV "can fit fairly" the experimental data for Au+Au collisions at $\sqrt{s_{NN}}$ = 3 GeV, whereas the momentum-independent nuclear mean fields (MID) with $K_0$ = 230 MeV and $K_0$ = 380 MeV "can only partially describe" the experimental results. We quantify these statements by performing $\chi^2$ goodness-of-fit tests for the comparison between the theoretical calculations and the experimental distributions presented in that work. For the complete set of data, all three calculations show highly statistically significant disagreement with the experimental data, with ($p$-value) $p < 0.001$. For the 0–10% centrality class, however, the MDI calculation with $K_0$ = 230 MeV gives $p = 0.49$, indicating no statistically significant disagreement with the experimental data for the proton transverse-momentum distributions and the rapidity dependence of the mean transverse momentum, whereas both MID calculations give $p < 0.001$. For the 10–40% centrality class, the analysis of the directed flows of protons, $K^+$ mesons, and $\Lambda$-hyperons gives the largest p-values for MDI with $K_0$ = 230 MeV and MID with $K_0$ = 380 MeV, with $p = 1.00$ and $p = 0.97$, respectively. In contrast, the analysis of the corresponding elliptic flow data gives $p < 0.001$ for all three calculations. Our results demonstrate that the qualitative statements "fit fairly" and "only partially describe" depend substantially on the particular observables and centrality classes considered and cannot by themselves establish statistically acceptable agreement between the calculations and the experimental data.

To study the properties of nuclear matter at extremely high baryon densities and low temperatures, the STAR Collaboration performed measurements of proton transverse-momentum $p_T$ spectra in central (0–10%) Au+Au collisions and of directed and elliptic flows of identified hadrons ($p$, $K^+$, and $\Lambda$) in semi-central (10–40%) Au+Au collisions at $\sqrt{s_{NN}}$ = 3 GeV [1, 2]. The transverse-momentum spectra provide information on the collective dynamics, thermodynamic properties, and hadronization mechanisms of the nuclear medium [3, 4]. Flow observables provide important information on the nuclear equation of state (EoS), possible phase transitions to QGP matter, and hadronic mean-field potentials at high baryon densities [5–7]. These STAR data were compared in [8] with calculations using a momentum-dependent nuclear mean field (MDI) with incompressibility $K_0$ = 230 MeV (MDI 230) and momentum-independent nuclear mean fields (MID) with $K_0$ = 230 MeV and $K_0$ = 380 MeV (MID 230 and MID 380). The authors of [8] stated that the MDI calculation provides a "fair fit" to the experimental data, whereas both MID calculations "only partially reproduce" the experimental results. Since the terms "fair" and "partially" do not provide a statistically well-defined measure of the agreement between theory and experiment, we perform a $\chi^2$ goodness-of-fit test and calculate $p$-values to provide a quantitative statistical assessment of the agreement for the three calculations.

We carried out the calculations using the following scheme. Suppose that a physical observable $y$ is experimentally measured and theoretically calculated at $N_1$ kinematical points, with corresponding values and experimental and theoretical standard uncertainties:

v.kizka@ispnpp.kiev.ua
valeriy.kizka@karazin.ua

$y_1^{\exp} + \sigma_1^{\exp}, ..., y_{N_1}^{\exp} + \sigma_{N_1}^{\exp}$ and $y_1^{th} + \sigma_1^{th}, ..., y_{N_1}^{th} + \sigma_{N_1}^{th}$, respectively. The experimental data points are treated as statistically independent. For a point-by-point comparison of experimental and theoretical results, we calculate $\chi^2$ statistics as:

$$\chi^2_{obs_1} = \sum_{i=1}^{N_1} \frac{(y_i^{\exp} - y_i^{th})^2}{(\sigma_i^{\exp})^2 + (\sigma_i^{th})^2}. \qquad (1)$$

When no theoretical uncertainty is provided, we set $\sigma_i^{th} = 0$. If no model parameters are fitted to the data, the number of degrees of freedom $\nu$ is equal to the number of data points $\nu = N_1$. The resulting value is referred to as the observed $\chi^2$ value, $\chi^2_{obs_1}$. We then calculate the right-tail probability of the $\chi^2$ distribution, i.e., the $p$-value:

$$p = P(\chi^2_{N_1} \geq \chi^2_{obs_1}) = 1 - F_{\chi^2}(\chi^2_{obs_1}; N_1) = \int_{\chi^2_{obs_1}}^{\infty} \frac{t^{\frac{N_1}{2}-1} e^{-t/2}}{2^{N_1/2}\Gamma\left(\frac{N_1}{2}\right)} dt, \qquad (2)$$

which represents the probability, under the null hypothesis that the theoretical calculation is consistent with the experimental data, of obtaining a $\chi^2$ value at least as large as the observed one. If $p < 0.05$, the discrepancy between theory and experiment is considered statistically significant and the null hypothesis is rejected at the 5% significance level. For $p < 0.01$ and $p < 0.001$, the discrepancy is considered very statistically significant and highly statistically significant, respectively, corresponding to significance levels of 1% and 0.1%. We used the standard TMath::Prob($\chi^2_{obs}$, $\nu$) function of the C++-based ROOT 5.22 framework developed at CERN [9] to calculate the $p$-value.

If an additional observable $z$ is measured and calculated for $N_2$ kinematical points, its contribution is calculated analogously, yielding $\chi^2_{obs_2}$. Assuming that the two sets of data are statistically independent, the $\chi^2$ values can be summed:

$$\chi^2_{obs_{1,2}} = \chi^2_{obs_1} + \chi^2_{obs_2}, \quad N = N_1 + N_2. \qquad (3)$$

The $p$-value for the complete set of $N$ data points is then obtained from:

$$p = P(\chi^2_N \geq \chi^2_{obs_{1,2}}), \qquad (4)$$

with $N$ degrees of freedom. In the case of $m$ independent observables, we sum all $m$ $\chi^2$ contributions and the corresponding numbers of degrees of freedom:

$$\chi^2_{obs_{1-m}} = \sum_{j=1}^{m} \chi^2_{obs_j}, \quad N = \sum_{j=1}^{m} N_j, \quad p = P(\chi^2_N \geq \chi^2_{obs_{1-m}}), \qquad (5)$$

where $N_j$ is the number of data points for the $j$-th observable.

We took the data from Figs. 3 and 4 of [8] to obtain *p*-values for the comparison of the three nuclear mean-field calculations with the experimental data for central collisions (0–10%). In [8], proton $p_T$ spectra are presented in Fig. 3 for several rapidity intervals. There are 10 $p_T$ spectra corresponding to 10 rapidity intervals, with a total of 176 experimental data points. The theoretical results are shown as lines with widths representing the theoretical uncertainties. For MID with $K_0$ = 230 MeV, the theoretical curve is shown for 174 of the 176 experimental points, while the last two experimental points have no corresponding theoretical values. Figure 4 contains 20 experimental points, while the theoretical results are shown by lines with markers. The meaning of the markers is not specified in [8]. Table 1 presents the $\chi^2$ values and *p*-values for the comparison of the three theoretical calculations with the experimental distributions shown in Figs. 3 and 4 of [8]. As seen, only MDI with $K_0$ = 230 MeV shows no statistically significant disagreement for the complete set of 0 – 10% data, with $p$ = 0.49. MID with $K_0$ = 230 MeV also shows no statistically significant disagreement with the experimental proton $p_T$ spectra, with $p$ = 0.87. Thus, only two calculations, MDI 230 and MID 230, are statistically consistent with the measured proton $p_T$ spectra, which differs from the qualitative assessment in [8] that all three calculations "fairly reproduce" the experimental data. All three calculations show highly statistically significant disagreement with the experimental rapidity dependence of the mean transverse momentum of protons, with $p < 0.001$ in all cases. Thus, for central Au+Au collisions at $\sqrt{s_{NN}}$ = 3 GeV, our results do not support the qualitative assessment in [8] that MDI 230 and MID 380 "reasonably fit" this observable.

***Table 1.*** *Calculated $\chi^2$ and p-values for the comparison of experimental and theoretical proton $p_T$ spectra and the rapidity dependence of the mean transverse momentum in central (0 – 10%) Au+Au collisions at $\sqrt{s_{NN}}$ = 3 GeV, using the data from Figs. 3 and 4 of [8]. The last two columns show the total $\chi^2$ values and p-values for the 0 – 10% centrality class, calculated using Eqs. 3 and 4. p-values indicating no statistically significant disagreement between theory and experiment are highlighted in bold.*

| Figure in [8] | model | $\chi^2_{obs_1} / N_1$ | *p*-value | Figure in [8] | model | $\chi^2_{obs_2} / N_2$ | *p*-value | $\chi^2_{obs_{1,2}} / N_1 + N_2$ | *p*-value |
|---|---|---|---|---|---|---|---|---|---|
| Figure 3, $p_T$-spectra of protons | MDI 230 | 133.07/176 | **0.99** | Figure 4, $\langle p_T \rangle(y)$, protons | MDI 230 | 62.97/20 | $2.45\cdot10^{-6}$ | 196.04/196 | **0.49** |
| | MID 230 | 152.93/174 | **0.87** | | MID 230 | 269.10/20 | $1.57\cdot10^{-45}$ | 422.03/194 | $8.45\cdot10^{-19}$ |
| | MID 380 | 276.81/176 | $1.84\cdot10^{-6}$ | | MID 380 | 80.10/20 | $3.77\cdot10^{-9}$ | 356.91/196 | $1.77\cdot10^{-11}$ |

Using the data from Figs. 5 – 7 of [8], we calculated the p-values for the 10-40% centrality class for the directed and elliptic flows of protons, $K^+$ mesons, and associated $\Lambda$ hyperons. The results are presented in Table 2. As seen from the last two rows of the table, all three models show highly statistically significant disagreement with the experimental data for the complete set of observables in this centrality class, with all $p < 0.001$. At the same time, MDI 230 and MID 380 have no statistically significant disagreement with experimentally measured directed flows of identified hadrons, with all p-values close to or equal to unity. In contrast, all three calculations show highly statistically significant disagreement with the experimental elliptic flows, with $p < 0.001$, with the exception of MDI 230 for the elliptic flow of $\Lambda$ hyperons, for which $p$ = 0.12 (no statistically significant disagreement).

Table 3 presents the final results, including the total $\chi^2$ values and $p$-values obtained by combining the central and semi-central data for the comparison of the three calculations with the experimental data. We see that all three models show highly statistically significant disagreement with the experimental data, with $p < 0.001$ in all cases. This contradicts the conclusion based on the qualitative assessment in [8] that MDI 230 provides a "fair fit" to the experimental data.

***Table 2.*** *Calculated $\chi^2$ and p-values for the comparison of experimental and theoretical directed and elliptic flows of protons, $K^+$ mesons, and $\Lambda$ hyperons in semi-central (10–40%) Au+Au collisions at $\sqrt{s_{NN}}$ = 3 GeV, using the data from Figs. 5 – 7 of [8]. The last two rows show the total $\chi^2$ values and p-values for the 10–40% centrality class, calculated using Eq. (5). p-values indicating no statistically significant disagreement between theory and experiment are highlighted in bold.*

| Figure in [8] | Fig.5a, $\upsilon_1(y)$, $p$ | | | | Fig.5b, $\upsilon_2(y)$, $p$ | | |
|---|---|---|---|---|---|---|---|
| model | MDI 230 | MID 230 | MID 380 | model | MDI 230 | MID 230 | MID 380 |
| $\chi^2_{obs_3} / N_3$ | 4.18/12 | 59.31/12 | 4.40/12 | $\chi^2_{obs_6} / N_6$ | 30.92/12 | 125.65/12 | 205.90/12 |
| $p$-value | **0.98** | $3.02\cdot10^{-8}$ | **0.97** | $p$-value | $2.02\cdot10^{-3}$ | $4.60\cdot10^{-21}$ | $3.86\cdot10^{-17}$ |
| Figure in [8] | Fig.7a, $\upsilon_1(y)$, $K^+$ | | | | Fig.7c, $\upsilon_2(y)$, $K^+$ | | |
| $\chi^2_{obs_4} / N_4$ | 9.31/20 | 229.79/20 | 28.55/20 | $\chi^2_{obs_7} / N_7$ | 456.31/20 | 192.41/20 | 249.00/20 |
| $p$-value | **0.98** | $1.32\cdot10^{-37}$ | **0.97** | $p$-value | $3.94\cdot10^{-84}$ | $3.55\cdot10^{-30}$ | $1.82\cdot10^{-41}$ |
| Figure in [8] | Fig.7b, $\upsilon_1(y)$, $\Lambda$ | | | | Fig.7d, $\upsilon_2(y)$, $\Lambda$ | | |
| $\chi^2_{obs_5} / N_5$ | 12.47/20 | 27.44/20 | 1.15/20 | $\chi^2_{obs_8} / N_8$ | 10.08/6 | 26.38/6 | 27.07/6 |
| $p$-value | **0.90** | **0.12** | **1.00** | $p$-value | **0.12** | $1.89\cdot10^{-4}$ | $1.40\cdot10^{-4}$ |
| $\chi^2_{obs_{3-5}} / N_3 + N_4 + N_5$ | 25.96/52 | 316.54/52 | 34.10/52 | $\chi^2_{obs_{6-8}} / N_6 + N_7 + N_8$ | 497.31/38 | 344.44/38 | 381.97/38 |
| $p$-value | **1.00** | $1.36\cdot10^{-39}$ | **0.97** | $p$-value | $2.28\cdot10^{-81}$ | $4.97\cdot10^{-51}$ | $2.24\cdot10^{-58}$ |
| $\chi^2_{obs_{3-8}} / N_3 + N_4 + N_5 + N_6 + N_7 + N_8$ | 523.27/90 | 660.98/90 | 416.07/90 | | | | |
| $p$-value | $2.55\cdot10^{-62}$ | $8.92\cdot10^{-88}$ | $2.12\cdot10^{-43}$ | | | | |

We may conclude that only part of the experimental data is statistically consistent with each model. MDI 230 is statistically consistent with the largest fraction of the experimental data, followed by MID 380. Thus, the statements that a model "fairly" or "partially" reproduces the experimental data depend on the particular observable and centrality class considered and do not provide a statistically adequate quantitative assessment of the agreement between the model and the experimental data.

***Table 3.*** *Total $\chi^2$ values and p-values calculated using Eq. (5) for the comparison of experimental and theoretical proton $p_T$ spectra and the rapidity dependence of the mean transverse momentum in central (0–10%) Au+Au collisions, and directed and elliptic flows of protons, $K^+$ mesons, and $\Lambda$ hyperons in semi-central (10–40%) Au+Au collisions at $\sqrt{s_{NN}}$ = 3 GeV, using the data from Figs. 3–7 of [8].*

| | MDI 230 | MID 230 | MID 380 |
|---|---|---|---|
| $\chi^2_{obs_{1-8}} / N_1 + N_2 + N_3 + N_4 + N_5 + N_6 + N_7 + N_8$ | 719.31 / 286 | 1083.01 / 284 | 772.98 / 286 |
| $p$-value | $3.36\cdot10^{-39}$ | $2.56\cdot10^{-93}$ | $1.95\cdot10^{-46}$ |